# High-resolution single-molecule mass measurement of megadalton assemblies in solution

Tereza Roesel[1]†, Evangelos Efraimidis[1,2]†, Niklas Hansen[3,4], Yulia Yancheva[5,1], Miroslav Hekrdla[1], Katarzyna M. Tych[5,6], Vladimíra Petráková[3], Barbora Špačková[1]*

[1] FZU - Institute of Physics of the Czech Academy of Sciences, Na Slovance 1999/2, 182 21 Prague, Czechia
[2] Faculty of Mathematics and Physics, Charles University, Ke Karlovu 3, 121 16 Prague 2, Czech Republic
[3] J. Heyrovský Institute of Physical Chemistry, Czech Academy of Sciences, Dolejškova 3, 182 23 Prague, Czechia
[4] Department of Physical Chemistry, Faculty of Chemical Engineering, University of Chemistry and Technology, Technická 5, 166 28 Prague 6, Czechia
[5] Chemical Biology 1, University of Groningen, Nijenborgh 7, 9747 AG Groningen, The Netherlands
[6] Institute for Microsystems Engineering (IMTEK), Faculty of Engineering, University of Freiburg, Germany

†Co-first authors; these authors contributed equally.
*Corresponding author

# Abstract

Resolving heterogeneity in megadalton assemblies requires precise single-molecule mass measurements in solution. Mass photometry infers mass from individual molecular surface-landing events, and event-to-event measurement variability limits precision. Nanofluidic scattering microscopy overcomes this limitation by continuously tracking molecules in motion, enabling repeated sampling and temporal averaging of these fluctuations. Benchmarking with 4.5 MDa DNA origami demonstrates up to a fourfold improved resolution, approaching the performance of ensemble-averaged native mass spectrometry.

# Main

Precise determination of the molecular mass of biomolecules under native conditions is essential for understanding their composition, structure, and function. Even subtle mass differences can reflect changes in stoichiometry, subunit exchange, or ligand binding, often with profound functional or pathological consequences. Resolving such closely related molecular species requires high mass resolution. Native mass spectrometry (MS) has become the benchmark technique for high-resolution mass analysis of intact proteins and protein complexes.[1] However, because the measured signal arises from ensembles of thousands to millions of molecules, the analysis of large and compositionally heterogeneous assemblies remains challenging [2]. In addition, native MS requires transfer of biomolecular assemblies from solution into the gas phase, where ionization and desolvation can perturb fragile complexes despite carefully optimized conditions [3].

Optical interferometric methods offer a complementary route to biomolecular mass measurement closer to physiological conditions, as they allow measurements directly in solution with minimal constraints on its composition [4]. By detecting individual biomolecules, these methods can also resolve sample heterogeneity and reveal rare or coexisting species that are difficult to distinguish in ensemble measurements. However, the most widespread implementation, mass photometry (MP) [5], provides substantially lower mass resolution than native MS—by more than an order of magnitude in the megadalton regime [6].

We show that nanofluidic scattering microscopy (NSM) overcomes a major constraint that limits the resolution of optical single-molecule mass measurement. Instead of detecting single surface-landing events, NSM continuously tracks individual biomolecules as they flow in solution. We illustrate this capability by measuring 4.5 MDa DNA origami with up to fourfold higher resolution than commercial MP.

In optical interferometric methods, individual biomolecules are detected through interference between the field scattered by the molecule, $E_s$, and a coherent reference field, $E_r$, originating from the same illumination via reflection or scattering at an interface (Fig. 1a). The detected intensity is

$$I_{det} = I_s + I_r + 2\sqrt{I_s I_r} cos\varphi, \quad (1)$$

where $I_r = |E_r|^2$, $I_s = |E_s|^2$ are the reference and scattered field intensities, and $\varphi$ is the relative phase between the two fields. The molecular contribution is extracted by removing the dominant reference field intensity, typically estimated by temporal referencing from frames acquired at the same location in the absence of the molecule. The resulting interferometric contrast,

$$C = \frac{I_{det} - I_r}{I_r} \quad (2)$$

reduces for weak scatterers ($I_s \ll I_r$) to

$$C \approx 2|E_s| cos\varphi / |E_r|. \quad (3)$$

In the Rayleigh regime, $|E_s|$ scales linearly with molecular polarizability. For biomolecules of similar composition, polarizability is approximately linearly proportional to mass (Fig. 1b) [5,7], enabling quantitative mass determination after calibration with molecular standards.

Uncertainty in the measured contrast—and thus in the inferred molecular mass—arises from (i) position-dependent variations, (ii) orientation-dependent variations, and (iii) light-dependent intensity noise, including shot noise and instrumental drifts. Position-dependent variations result from spatial heterogeneity in the reference field amplitude and phase caused by substrate roughness and residual surface scatterers (e.g., adsorbed molecules) [8] (Fig. 1c). Orientation-dependent variations result from anisotropic molecular scattering, where the measured contrast depends on molecular orientation relative to the illumination and detection geometry [9] (Fig. 1c).

In MP, molecules are detected upon landing on the surface. The molecular contrast is extracted by differential imaging, in which the intensity acquired before the landing event serves as the reference for the subsequent frame containing the molecule (Fig. 1d). Increasing the integration time reduces photon shot noise, but only up to a characteristic optimum beyond which drifts dominate and noise increases [5] (Fig. 1d inset). Since individual molecules arrive at different surface locations and with different orientations, both reference-field heterogeneity and orientation-dependent scattering significantly contribute to variability in the measured molecular contrast and inferred mass.

NSM overcomes key limitations of MP by repeatedly sampling the same molecule as it transits through the illuminated nanofluidic channel (Fig. 1e), where scattering from the channel provides the reference field [7]. NSM acquires many independent measurements of the molecular contrast, while maintaining a short integration time for each frame and independently estimating the reference field intensity (Fig. 1f). The effective observation time can thus far exceed the optimal single-frame integration time used in MP, enabling more effective suppression of light-dependent intensity noise. In addition, as each molecule traverses different positions and orientations (Fig. 1g), variability in contrast that acts as a fixed offset in single surface-landing measurements becomes a time-dependent fluctuation that is reduced by averaging along the molecular trajectory. Consequently, the standard deviation of the trajectory-averaged contrast decreases as

$$\sigma_C(t_{obs}) = \frac{\sigma_C(t_{int})}{\sqrt{N}}, \quad (4)$$

where $\sigma_C(t_{int})$ is the per-frame contrast standard deviation for integration time $t_{int}$, and $N$ is the number of independent measurements acquired during the observation time ($t_{obs}$). Because molecular mass is directly proportional to molecular contrast, the uncertainty in the measured mass decreases accordingly. As a result, longer observation times are expected to produce progressively narrower single-species mass distributions (Fig. 1h), reflected by a reduced full width at half maximum (FWHM) centered at mass $m$ and correspondingly higher mass resolution, typically defined as $R = m/FWHM$.

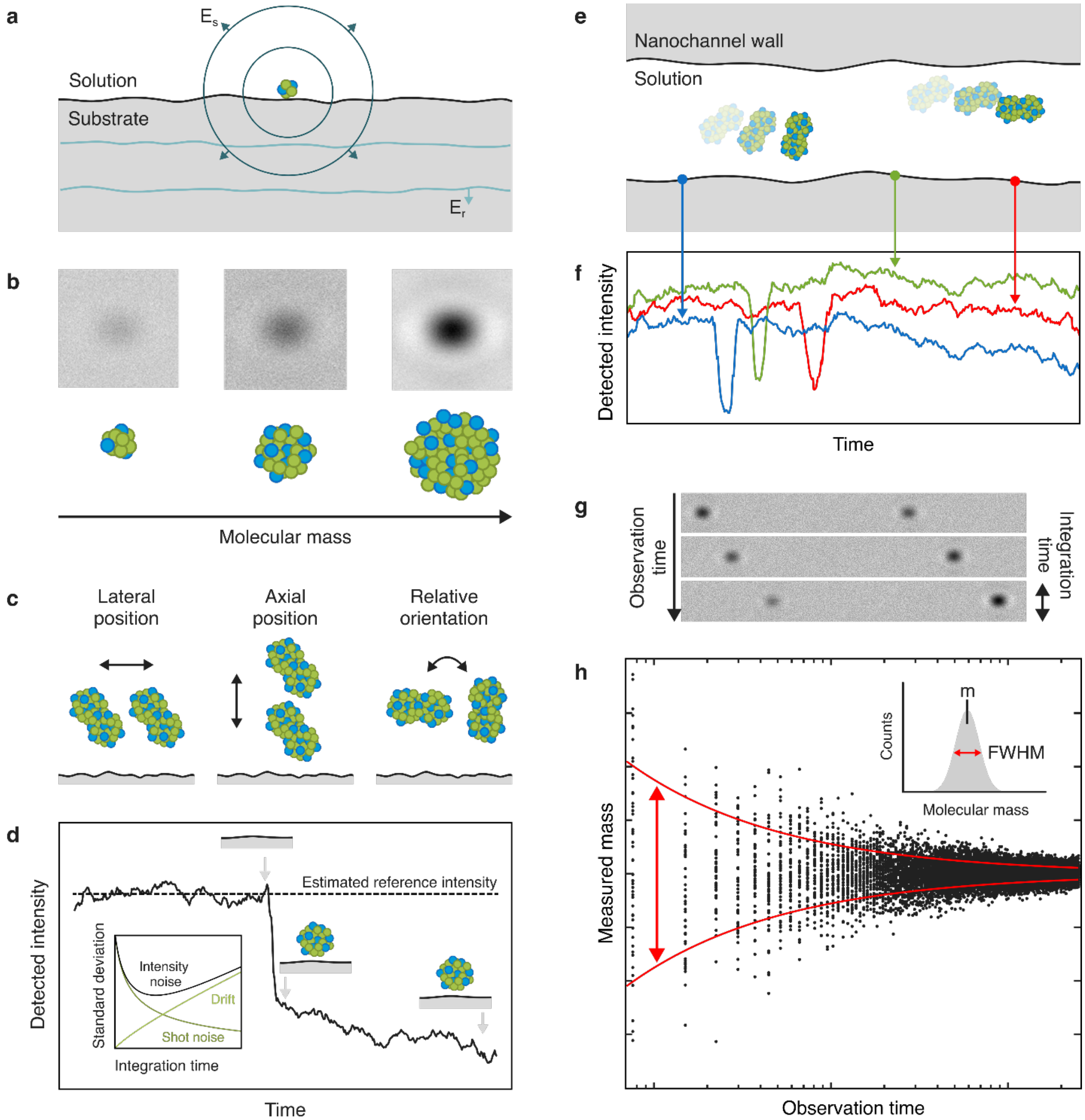

**Figure 1. Interferometric signal, sources of uncertainty, and principle of averaging-based mass resolution improvement. (a)** Principle of optical interference mass measurement. Light scattered from a molecule ($E_s$) interferes with the reference field ($E_r$). **(b)** Visualization of interferometric contrast images for molecules of increasing mass. Optical contrast scales proportionally with molecular mass. **(c)** Spatial heterogeneity of the reference field amplitude and phase leads to position-dependent variations in molecular contrast. Scattering of anisotropic particles leads to orientation-dependent variations in molecular contrast. **(d)** Representative time trace of detected intensity at a fixed position in MP. Fluctuations arise from intensity noise, whereas the abrupt intensity drop corresponds to a molecular landing event. Inset: Schematic dependence of the contrast standard deviation arising from light-dependent intensity noise on integration time. Photon shot noise decreases with increasing integration time, whereas instrumental drifts typically increase. **(e)** Schematic of NSM detection. A biomolecule is transported through an illuminated nanofluidic channel. **(f)** Representative time trace of detected intensity at three fixed positions in NSM. Fluctuations arise from intensity noise, whereas transient drops in intensity correspond to passing molecules. **(g)** Visualization of interferometric contrast images showing two molecules passing through the nanochannel at different times. **(h)** Expected dependence of single-molecule mass precision in NSM on observation time. Each dot represents the trajectory-averaged mass of an individual molecule. Longer observation times reduce the spread around the nominal mass value, leading to narrower mass distributions and decreased FWHM (red).

To benchmark NSM in the megadalton regime, we used DNA origami at a concentration of 60 nM (Methods) — a stable, monodisperse model system with a nominal mass of 4.5 MDa. Individual DNA origami were imaged under different flow rates within the same nanofluidic channel (cross-section 47 × 206 $nm^2$). Thereby, we were able to control the observation time and, correspondingly, the number of independent mass measurements per molecule. Representative interferometric image sequences of DNA origami transiting through the nanochannel are shown in Extended Data Movie 1. Figures 2a,b show kymographs and mass distributions of the fastest and the slowest flow rates, respectively. The full dataset spanning velocities of 96, 63, 35, and 14 μm/s, corresponding to observation times of 0.3, 0.5, 0.9, and 2.2 s, is provided in SI Fig. 1a,b.

For comparison, the same sample was measured by commercial MP (TwoMP, Refeyn; Methods). To ensure reliable identification of individual landing events, measurements were performed at different concentrations. At the optimized concentration (370 pM), differential images and the corresponding mass distribution are shown in Fig. 2c.

In NSM, reducing the flow velocity increased the observation time and, consistent with Eq. 2, led to a systematic narrowing of the mass distributions down to ~71 kDa (Fig. 2d). The resolution thus increased (Fig. 2e) up to $R$ = 63. The achieved performance places optical single-molecule mass measurements at approximately two- to threefold of ensemble-averaged native MS ($R \approx$ 100 - 200 in a similar mass range [6]). In contrast, MP yielded substantially broader peaks, with a minimum FWHM of 301 kDa (R = 15) at optimized concentration, corresponding to approximately fourfold lower resolution than NSM. At higher concentration (3.7 nM), MP distribution broadened further (FWHM = 504 kDa, R = 8.9), as increased surface crowding caused partial overlap of molecular landing events and altered the local reference field (SI Fig. 1e). The MP histograms also exhibit an additional pronounced signal population below ~2 MDa, likely arising from partial lateral motion of DNA origami on the surface after landing. Such low-contrast surface-associated events would obscure genuine low-mass species, highlighting a key limitation of surface-based measurements for heterogeneous samples spanning broad mass ranges.

The improved resolving power of NSM revealed an additional population at an inferred mass of ~5 MDa (Fig. 2a,b) that appeared only as a partially resolved shoulder in MP measurements (Fig. 2c). AFM imaging showed that, in addition to the expected extended rod-like conformation, a subpopulation of DNA origami adopted more compact back-folded conformations (Extended Data Fig. 1). Although these structures contain the same amount of DNA, their altered geometry is expected to modify the scattering amplitude and therefore the interferometric contrast (Eq. 3). Consistent with this interpretation, finite-difference time-domain (FDTD) simulations predicted that back-folded conformations produce approximately 10% higher contrast compared to the extended conformation, matching the experimentally observed inferred mass (SI). This conformational heterogeneity also provides a likely explanation for the residual deviation of the experimental NSM resolution from the theoretical NSM limit (Methods), as it introduces genuine molecule-to-molecule variation in contrast that cannot be averaged away.

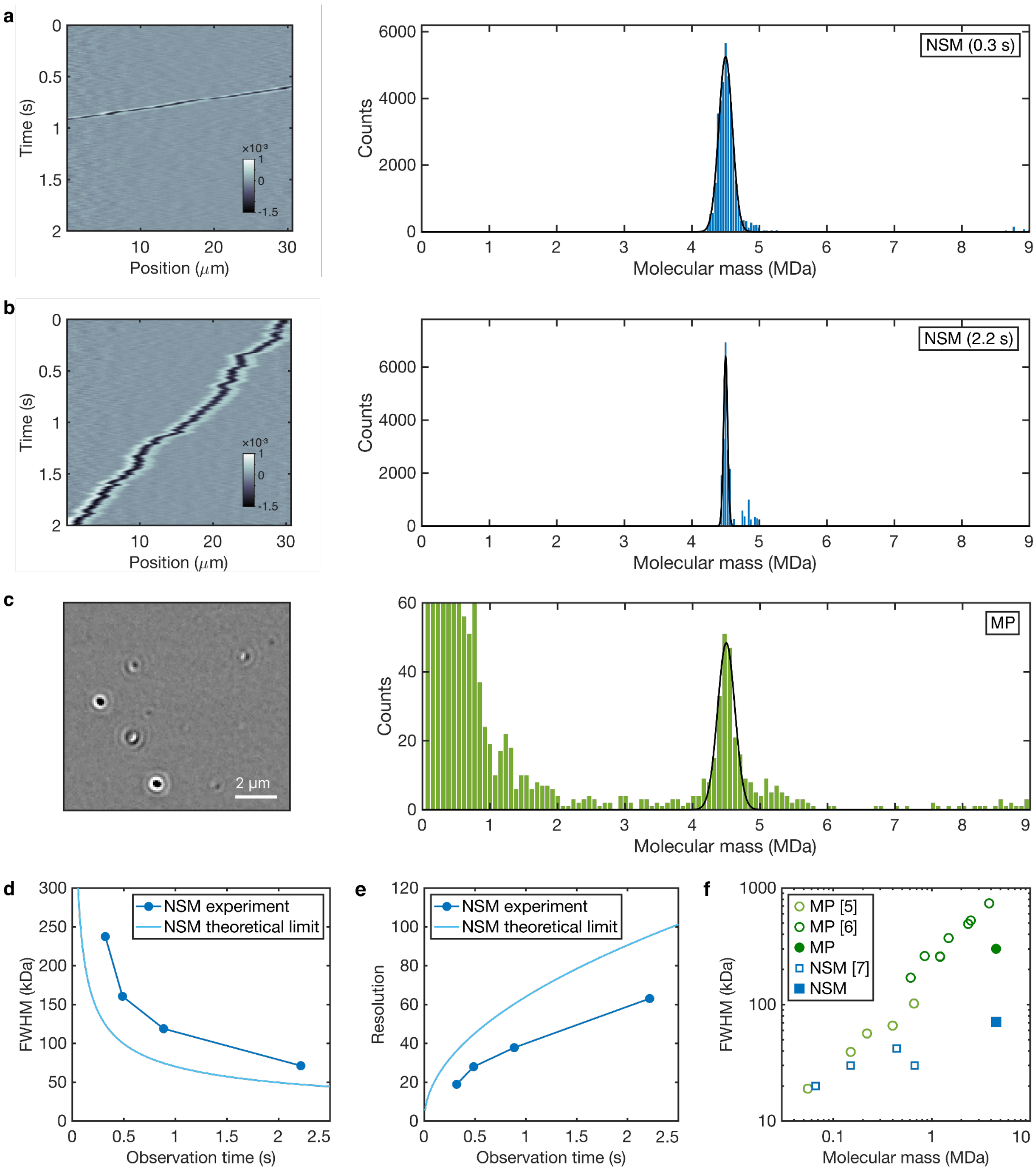

**Figure 2. Megadalton regime benchmarking with DNA origami. (a, b)** Representative NSM kymographs (left) of individual molecules tracked for 0.3 s and 2.2 s, respectively. Corresponding molecular mass histograms (right), in which each molecule contributes N counts according to the number of frames in the tracked trajectory. Black lines show Gaussian fits. **(c)** Representative MP differential image (left) of transient molecular adsorption events and the corresponding molecular mass distribution (right). For each of the histograms in (a-c), bin sizes were selected using Scott's rule (Methods). The black line shows a Gaussian fit. **(d, e)** Dependence of FWHM and mass resolution, respectively, on observation time for NSM, shown for both the experimental values and the calculated theoretical limit (Methods). **(f)** Dependence of FWHM on molecular mass. Literature values for NSM and MP [5–7] are plotted alongside the experimentally determined NSM and MP values from this work, highlighting the improved performance of NSM in the megadalton regime.

We next compared our measurements with previously reported NSM and MP data [5–7] by plotting the FWHM of the mass distribution as a function of molecular mass (Fig. 2f). The MP measurement reported here outperforms previously published MP data at a similar mass range, likely reflecting advances in acquisition and data analysis. Notably, NSM shows increasingly lower FWHM than MP with increasing molecular mass. This favorable scaling indicates more effective suppression of mass-dependent noise and suggests that NSM will provide progressively higher mass resolution for larger biomolecular assemblies, including extracellular vesicles [10], synthetic nanocontainers [11], and partially assembled biomolecular condensates [12], for which precise single-particle mass measurements remain challenging.

In summary, NSM enables high-resolution single-molecule mass measurements by continuously tracking biomolecules in solution and averaging their interferometric contrast over extended trajectories. In the megadalton regime, this repeated-sampling strategy improves mass resolution by up to fourfold relative to commercial MP. This brings optical single-molecule mass measurements close to the resolving power of ensemble-averaged native MS while preserving continuous measurements of individual molecules directly in solution.

This level of mass precision enables discrimination of heterogeneous macromolecular assemblies differing by only tens of kilodaltons in subunit occupancy, ligand binding, or cargo loading. Such capability is particularly valuable for structurally dynamic systems, including ribosomes [13], viral capsids [14], and large multiprotein complexes [5], for which ensemble measurements obscure heterogeneity and existing optical methods often lack sufficient resolving power.

Importantly, the present performance is unlikely to represent the fundamental limit of NSM. Because mass precision improves with observation time, advances such as longer nanochannels, larger fields of view, parallelized imaging, and trapping-assisted measurements should further extend molecular trajectories while maintaining sufficient throughput, enabling substantially higher mass resolution.

More broadly, NSM establishes a new framework for molecular mass metrology. By combining continuous solution-phase measurements with high mass precision, it opens new opportunities to investigate dynamic state transitions of biomolecular systems under near-native conditions, that have remained inaccessible to existing single-molecule techniques.

# Methods

## DNA origami

Twelve-helix bundle DNA origami nanostructures (Extended Data Fig. 2) were designed using the caDNAno software [15]. The structures were synthesized using an M13mp18 single-stranded DNA scaffold (tilibit Nanosystems) and complementary staple strands (Metabion International AG) mixed at a 1:10 molar ratio in folding buffer (1× TAE, 15 mM $MgCl_2$). The mixture was incubated in a Biometra TAdvanced thermocycler (Jena Analytik) for ~25 hours following established protocols [16]. Excess staples were removed by ultrafiltration using 100 kDa Amicon MWCO Ultra centrifugal filters (Merck Millipore). Briefly, 100 µL of synthesis solution was diluted with 400 µL of folding buffer and centrifuged at 5,000 × g for 5 min. This washing step was repeated four times, each with 400 µL of fresh folding buffer. The concentrated origami was recovered by inverting the filter unit into a clean tube and centrifuging at 3,000 × g for 5 min. The final concentration was determined spectrophotometrically using an NP80 NanoPhotometer (Implen).

The M13mp18 scaffold contains 7,249 bases with the following composition: A = 1,766; T = 2,419; C = 1,530; G = 1,534. Using the molecular weights of each nucleotide, the scaffold mass was calculated to be 2.236 MDa. Upon hybridization with complementary staple strands to form the twelve-helix bundle—plus a small number of single-stranded overhangs at each rod end—the theoretical mass increases to 4.479 MDa, with an additional 29.2 kDa from extensions, yielding 4.508 MDa. Accounting for end-group chemistry of the staples (removing one phosphorus and two oxygens, and adding two hydrogens per staple), the corrected final molecular mass is 4.496 MDa.

## Fluidics and Nanofluidic chips

Nanofluidic chips were fabricated by ConScience AB (Gothenburg, Sweden) in the ISO Class 5–6 cleanroom facilities of Chalmers University of Technology. The devices were produced from 4-inch silicon wafers with a 2.1 µm thermally grown oxide layer and sealed to 175 µm-thick 4-inch Pyrex lids by fusion bonding. The fabrication process comprised direct lithography (electron-beam lithography for the nanochannels and laser lithography for the larger features), reactive-ion etching, fusion bonding, and dicing. Each chip contained arrays of nanochannels with a length of 53 µm and cross-sectional dimensions tailored in the range of 40–200 nm. For the experiments reported here, nanochannels with a cross-section of 47 × 206 nm² were used. These channels were connected to inlet and outlet ports via microchannels measuring 50 × 1.6 µm². The final channel geometries were verified by scanning electron microscopy (SEM, Extended Data Fig. 3). Sample transport through the nanochannels was driven by pressure-controlled flow using a microfluidic flow control system (Fluigent), with pressures up to 250 mbar applied to sample reservoirs containing 20 µl of sample.

## Optical setup

All optical measurements were performed using a custom-built dark-field microscope designed for high-sensitivity scattering detection (Extended Data Fig. 4). Illumination was provided by a polychromatic supercontinuum laser (FIU-15, NKT Photonics), spectrally filtered to a 500-610 nm range using the SuperK VARIA module. The filtered beam, with a total power of

~420 mW and a beam diameter of 1 mm, was first expanded by a beam expander (Broadband VIS 2X - 8X Research-Grade Variable Beam Expander, Edmund Optics) to a diameter of 4.5 mm and then directed by a D-shaped mirror into the back focal plane of a high numerical aperture oil immersion objective (AmScope, NA = 1.25). A second D-shaped mirror was placed in the detection path to block reflected light and enhance contrast. Scattered light from the sample was collected through the same objective and imaged onto a CMOS camera (MV4-D1280-L01-GT, Photonfocus AG) with a chip size of 8.45 × 6.76 mm. The final optical magnification was 100×. The nanofluidic chip was mounted into a custom-made chip holder positioned on the objective using a NanoMax three-axis precision stage (Thorlabs).

## Image acquisition configuration

The camera region of interest comprised 464 × 32 pixels positioned in the central sensor area. The exposure time was set to 20 µs, resulting in a frame rate of ~47,000 fps after accounting for the readout time. To reduce the data throughput to a rate compatible with the acquisition hardware, the camera's built-in FrameCombine function was employed to vertically concatenate every 207 consecutive frames into a single transmitted multiframe image, yielding an effective frame rate of ~142 fps (~7 ms per multiframe). Acquisition was performed in recordings of 2,000 multiframes (~14 s duration), after which the image data were transferred from the camera buffer. The multiframe images were subsequently averaged along the temporal axis (i.e. 207 frames) and the spatial axis corresponding to the width of a nanochannel image (i.e. 32 pixels), stored as uncompressed .tiff files, and the acquisition cycle was repeated. The time-averaging interval was selected to match the period of intensity fluctuations likely corresponding to the mechanical vibrations in the optical system, thereby directly suppressing this noise contribution during acquisition.

## NSM data processing

### Dark current correction

The CMOS signal includes an additive dark-current offset originating from thermally generated electrons in the sensor. To correct for temperature-dependent variations in this offset, dark frames were acquired under identical acquisition settings across the operating temperature range. For each acquisition, a temperature-matched dark frame was subtracted from the raw data prior to further analysis.

### Noise suppression

High-spatial-frequency noise in the interferometric images reduces the detectability of weak molecular signals. The dominant contributions arise from photon shot noise and spatially correlated CMOS detector noise, including residual fixed-pattern and row-correlated readout noise. To suppress these noise components while preserving molecular features, images were convolved with a 16-pixel moving-average kernel.

### Reference field intensity estimation and removal

Accurate estimation of the reference intensity $I_r$ is required to isolate the molecular contribution and calculate the interferometric contrast. The reference intensity in Eq. (1) can be expressed as

$$I_r(x,t) = I_0(x,t) \cdot |r(x)|^2, \quad (5)$$

where *r* is the component incorporating contributions from nanochannel polarizability, collection efficiency, and wavelength-dependent factors, while $I_0$ is the incident illumination intensity. In practice, $I_0$ exhibits both spatial variations arising from the Gaussian beam profile and wavefront distortions of the optical system and temporal variations arising from mechanical drift and light source intensity fluctuations. To separate these fluctuations from the rapidly varying molecular signal in the detected intensity, we apply a reference field intensity estimation procedure that combines spatial and temporal filtering.

First, a spatial moving-average filter with a kernel width of $2W_x + 1$ pixels was applied to the dark-current-corrected and noise-suppressed intensity data $I(x,t)$:

$$I_{r,x}(x,t) = mean\{I(i,t), i = x - W_x, \dots, x + W_x\}. \quad (6)$$

Second, a temporal moving-median filter with a window of $2W_t + 1$ frames was applied:

$$I_{r,t}(x,t) = median\left\{\frac{I(x,j)}{I_{r,x}(x,j)}, j = t - W_t, \dots, t + W_t\right\}. \quad (7)$$

The final reference intensity estimate was then obtained as

$$I_r\ (x,t) = I_{r,x}(x,t) \cdot I_{r,t}(x,t), \quad (8)$$

which captures both spatial and temporal variation of reference field intensity. The interferometric contrast kymograph was subsequently calculated as

$$C(x,t) = \frac{I(x,t) - I_r\ (x,t)}{I_r\ (x,t)}. \quad (9)$$

The filter parameters were empirically optimized using DNA origami monomers by minimizing contrast fluctuations along individual molecular trajectories, yielding $W_x = 25$ pixels and $W_t = 150$ frames.

**Particle tracking algorithm**

Single-particle tracking analysis was performed using sequential particle detection and trajectory linking steps. Particle detection identifies molecular signals above the background noise in individual frames, whereas trajectory linking associates detections across consecutive frames to reconstruct molecular trajectories.

Particle detection was performed by thresholding applied to interferometric contrast kymographs (Eq. 9), followed by local minima identification. The thresholding does not require local adjustment for background non-uniformity, since the background is removed during the preprocessing stage. The decision threshold was determined solely from the measured noise level and the selected false-positive probability [17], $P_{FP} = 10^{-5}$. Local minima were identified using the greyscale morphological operation of erosion with a structuring element matched to the expected molecular image size [18]. A structuring element width of 6 pixels was used.

Trajectory linking was performed using a two-step tracking algorithm [19]. In the first step, particles detected in consecutive frames were connected using a frame-by-frame linking procedure. This step was globally optimal in space and locally optimal in time. The frame-by-frame linking is formulated as a linear assignment problem that minimizes the sum of squared Euclidean distances, while accounting for the flow in the nanofluidic system. In the next step, the resulting shorter trajectory segments (tracklets) were connected by solving another linear assignment problem, approximating the computationally demanding global optimum in space and time. This approach is robust to short detection outages (e.g., due to signal fluctuations) and performs well under low particle-density conditions, which correspond to our experimental conditions. The implementation relies on the core MATLAB function *matchpairs*, with optional penalization of unpaired detections. Prior to the linking step, a gating operation discards candidate pairs with distances exceeding a user-defined cut-off threshold.

**Molecular interferometric contrast quantification**

Molecular interferometric contrast values were extracted from the interferometric contrast kymographs (Eq. 9) at positions determined by the particle detection and trajectory linking algorithm. To compensate for stochastic motion-induced broadening of the molecular point-spread function, contrast values were obtained by integrating the interferometric signal over the molecular image, with integration boundaries defined by the zero-crossings.

For each molecule, the interferometric contrast was calculated as the mean of the integrated contrast values evaluated along the trajectory. To correct for spatial variations in relative detection sensitivity arising from changes in the nanochannel scattering amplitude and phase across the field of view, the measured contrast values were normalized using an iterative separable decomposition procedure. The integrated contrast measured for the j-th molecule at spatial position $x_i$, denoted $C_j(x_i)$, was modeled as

$$C_j(x_i) \approx \overline{C_j} \cdot S(x_i), \quad (10)$$

where $\overline{C_j}$ is the molecular contrast corrected for the spatial dependent sensitivity and $S(x_i)$ is the relative spatial sensitivity profile.

In the first step of the iteration, the procedure was initialized by estimating $S(x_i) = 1$.

In the second step, using the estimation of $S(x_i)$, molecular contrasts were calculated as

$$\overline{C_j} = \frac{1}{N_j}\sum_{i=1}^{N_j} \frac{C_j(x_i)}{S(x_i)}, \quad (11)$$

where $N_j$ is the number of frames associated with molecule *j*.

In the third step, using the estimation of $\overline{C_j}$, a normalized sensitivity profile was calculated as

$$S(x_i) = \frac{1}{M_i}\sum_{j=1}^{M_i} \frac{C_j(x_i)}{\overline{C_j}}, \quad (12)$$

where $M_i$ is the number of measurements contributing to spatial position $x_i$.

Because the decomposition is scale-invariant, a normalization is imposed in the fourth step such that $\overline{S(x_i)} = 1$. Iterations between the second and fourth steps are repeated until convergence of the normalized contrast values, $C_j(x_i)/\underline{C_j}$. Convergence was defined when the relative change in the standard deviation of these normalized values between successive iterations was below $10^{-3}$.

This procedure effectively removes position-dependent instrumental biases, ensuring that the interferometric contrast measured for each trajectory accurately reflects the molecular mass, independent of the molecule's location along the nanochannel.

**Trajectory quality control and outlier rejection**

Detected molecules were subjected to automated quality control to remove artifactual trajectories. For each trajectory, we evaluated the number of observations *N*, the mean velocity, the initial and terminal longitudinal positions, and the standard deviation of integrated optical contrast calculated as,

$$\sigma_C = \sqrt{\frac{1}{N}\sum_{i=1}^{N}\left(\frac{C(x_i)}{S(x_i)} - \overline{C_J}\right)^2}, \ (13)$$

For each parameter, the distribution across all detected trajectories was characterized by its mean ($\mu$) and standard deviation ($\sigma$). A three-sigma criterion was applied to each parameter—trajectories exceeding $\mu + 3\sigma$ (or falling below $\mu - 3\sigma$, where applicable) were rejected. This procedure removes trajectories affected by optical instabilities, mislinked and truncated trajectories, and noise-induced false positives, while preserving the dominant single-molecule populations.

**Calibration to mass**

The molecular contrasts $\bar{C}$ were converted to mass via calibration against the nominal mass of DNA origami monomer. Mean contrast values and FWHM of the DNA origami populations were determined by Gaussian fitting of the molecular contrast histograms using Matlab *fit* function. Histogram bin widths were selected according to Scott's rule to provide statistically appropriate binning for Gaussian fitting. The contrast-to-mass calibration factor was calculated as $c = m_0/C_0$, where $m_0 = 4.5\ MDa$ is the nominal mass of DNA origami monomer and $C_0 = -0.94\ nm$ is the corresponding mean molecular contrast obtained from the Gaussian fit. Molecular masses of all detected molecules were then calculated as $m = c\bar{C}$. Importantly, the measured mean masses of monomeric, dimeric, and trimeric populations occurred at approximately integer multiples of the monomer mass, confirming the near-linear relationship between interferometric contrast and molecular mass (SI Fig. 2).

## Theoretical prediction of FWHM and resolution for NSM

To estimate the theoretical performance limit of NSM, we used the relationship between the trajectory-averaged contrast fluctuations and the number of independent measurements described in Eq. 4. Using the contrast-to-mass calibration factor *c*, the corresponding standard deviation in mass measurement is then given by

$$\sigma_m(t_{obs}) = c\,\frac{\sigma_C(t_{int})}{\sqrt{N}}. \quad (14)$$

To calculate the estimated $\sigma_m$ for molecules with a mass of 4.5 MDa, we have used $\sigma_C = 0.075\ nm$ corresponding to the mean per-frame contrast standard deviation of the monomeric DNA origami population measured at $t_{int} = 7\ ms$. Assuming Gaussian statistics, the expected FWHM of the mass distribution (Fig. 2d) was calculated as $FWHM = 2\sqrt{2ln2}\sigma_m$ and the corresponding resolution (Fig. 2e) was calculated as $R = 4.5\ MDa/FWHM$.

## Mass Photometry

The mass photometry measurements were performed on a Refeyn TwoMP instrument (Refeyn, UK) using AcquireMP software, provided by Refeyn. High-precision microscope cover glasses (24x50 mm, thickness 170 µm, Marienfeld, DE) were cleaned by immersing in 3% mucasol solution for at least 30 min. They were subsequently rinsed with 70% ethanol and then MilliQ water three times and dried using nitrogen gas. The slides were then coated with 0.01% poly-l-lysine (PLL). An 8 µL PLL droplet was pipetted on a clean cover glass, and another clean cover glass was put on top such that the long edges of the two cover glasses are perpendicular to each other, with the droplet in the middle, and ensuring that the PLL is evenly spread in the middle of the two cover glasses, for 30 s. The glasses were then separated, rinsed with MilliQ water, and dried with nitrogen gas. A six-well adhesive gasket (Refeyn, UK) was then placed in the middle of a clean, passivated cover glass, which was inserted into the instrument. All data were collected using the large-field-of-view (200 µm²) setting of the AcquireMP software.

After adding 10 µL of 1× TAE, 15 mM $MgCl_2$ buffer to one of the wells, the focus was found using the droplet-dilution setting. The sample was then diluted to concentrations of 7.4 nM and 740 pM and added to the well with buffer, yielding final concentrations of 3.7 nM and 370 pM. The frame integration time was 23 ms, and the duration of the measurement was 60 s.

Molecular contrast values were extracted from the sequence of differential images using the instrument software and converted to molecular mass via calibration against the nominal DNA origami monomer mass, analogous to the NSM analysis.

# Extended data

Extended Data Movie 1: Sequence of interferometric contrast images showing two DNA origami structures transiting through the nanochannel.

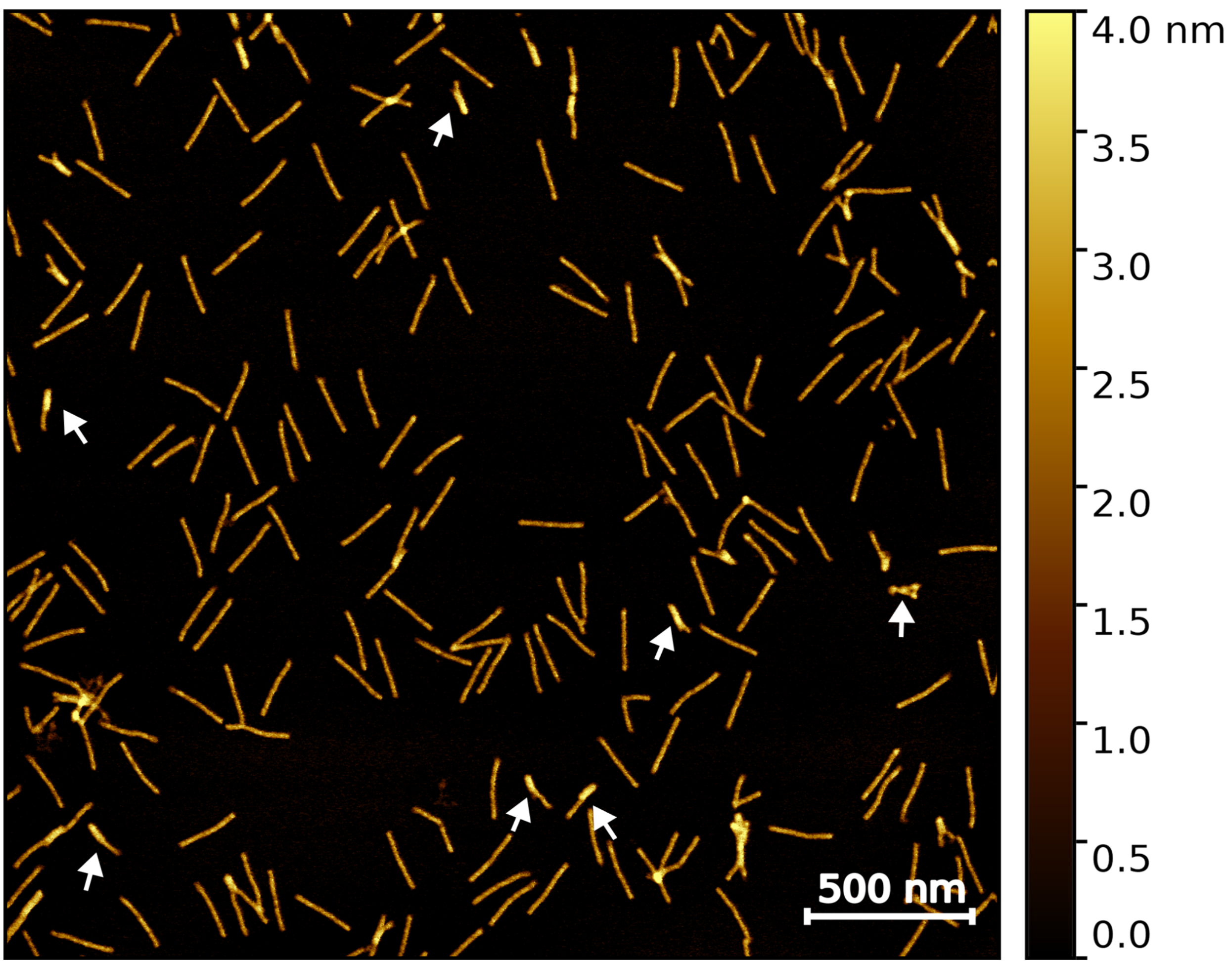

Extended Data Figure 1. AFM image of DNA origami rods. Representative AFM image showing predominantly correctly folded DNA origami rods, with arrows indicating a subpopulation of back-folded structures.

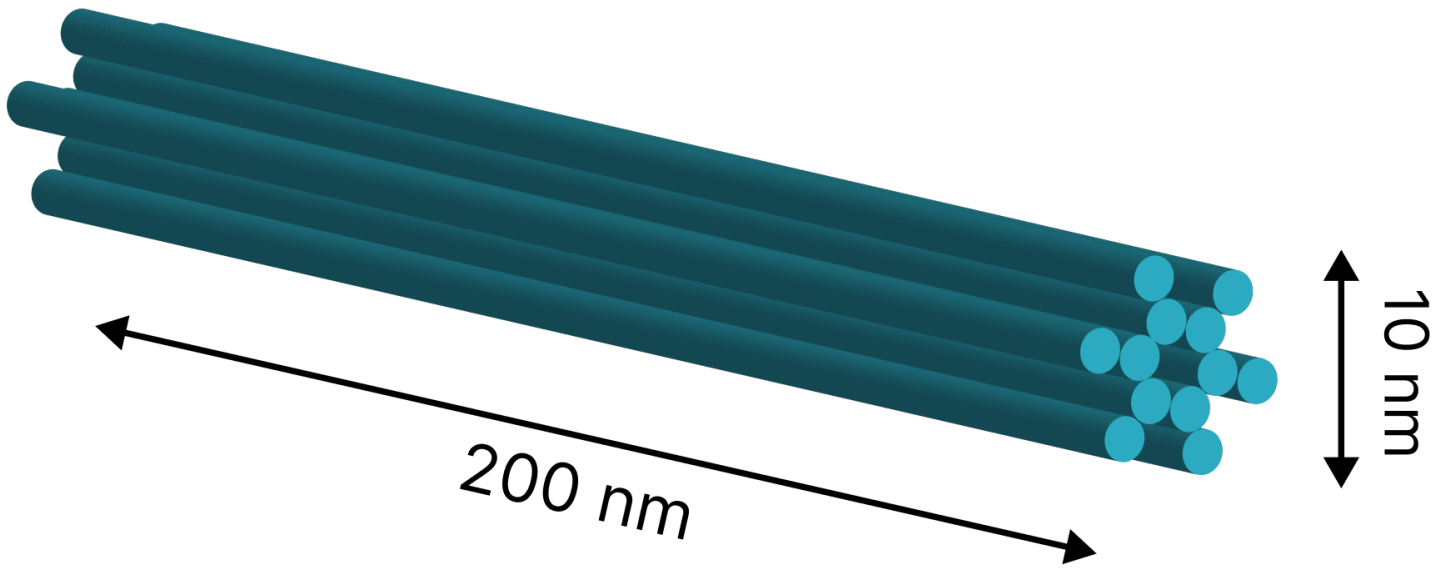

Extended Data Figure 2: Schematic of the DNA origami.

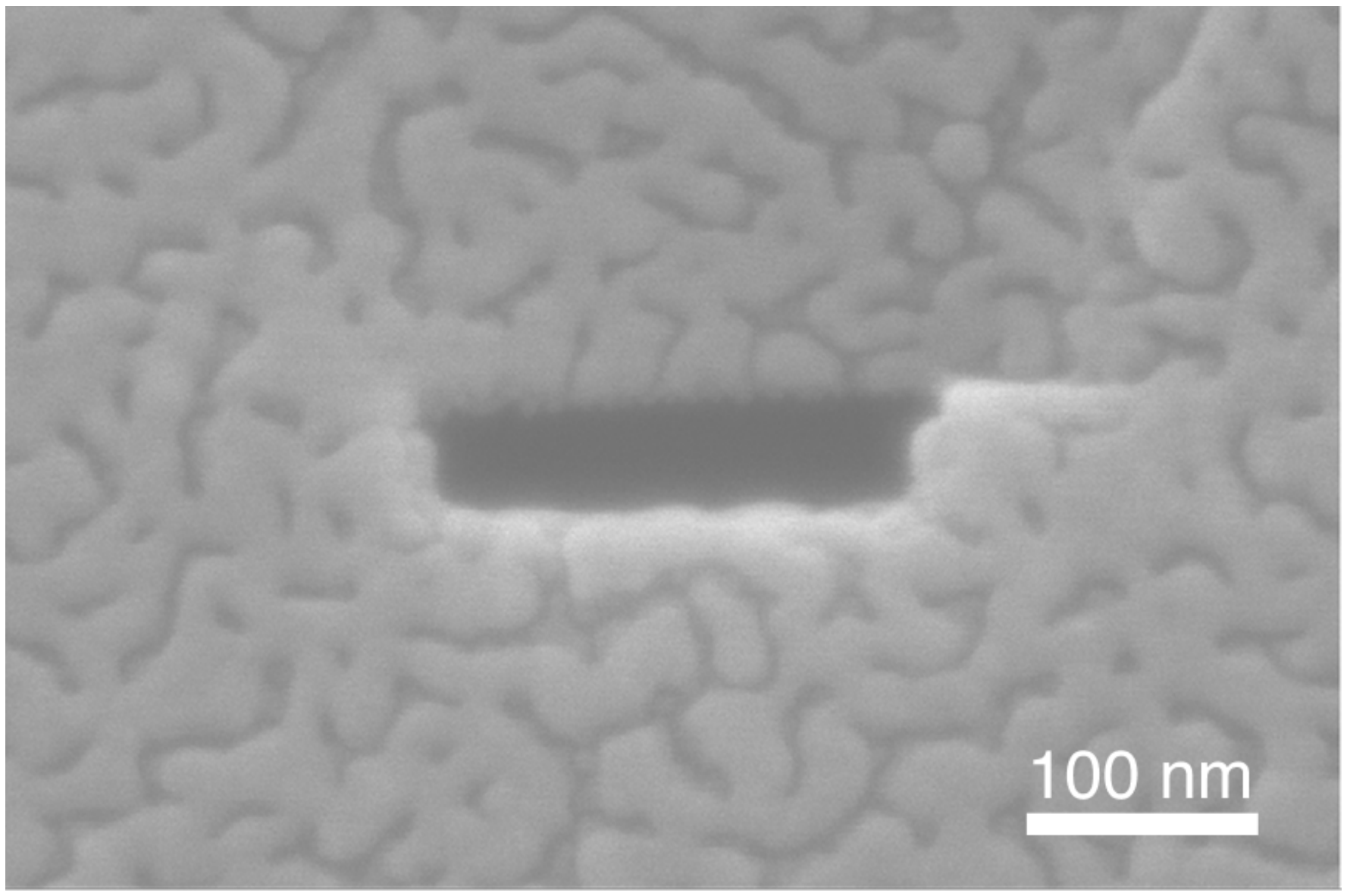

Extended Data Figure 3: Scanning electron microscopy images of the nanochannel with a cross-section of $47 \times 206$ nm².

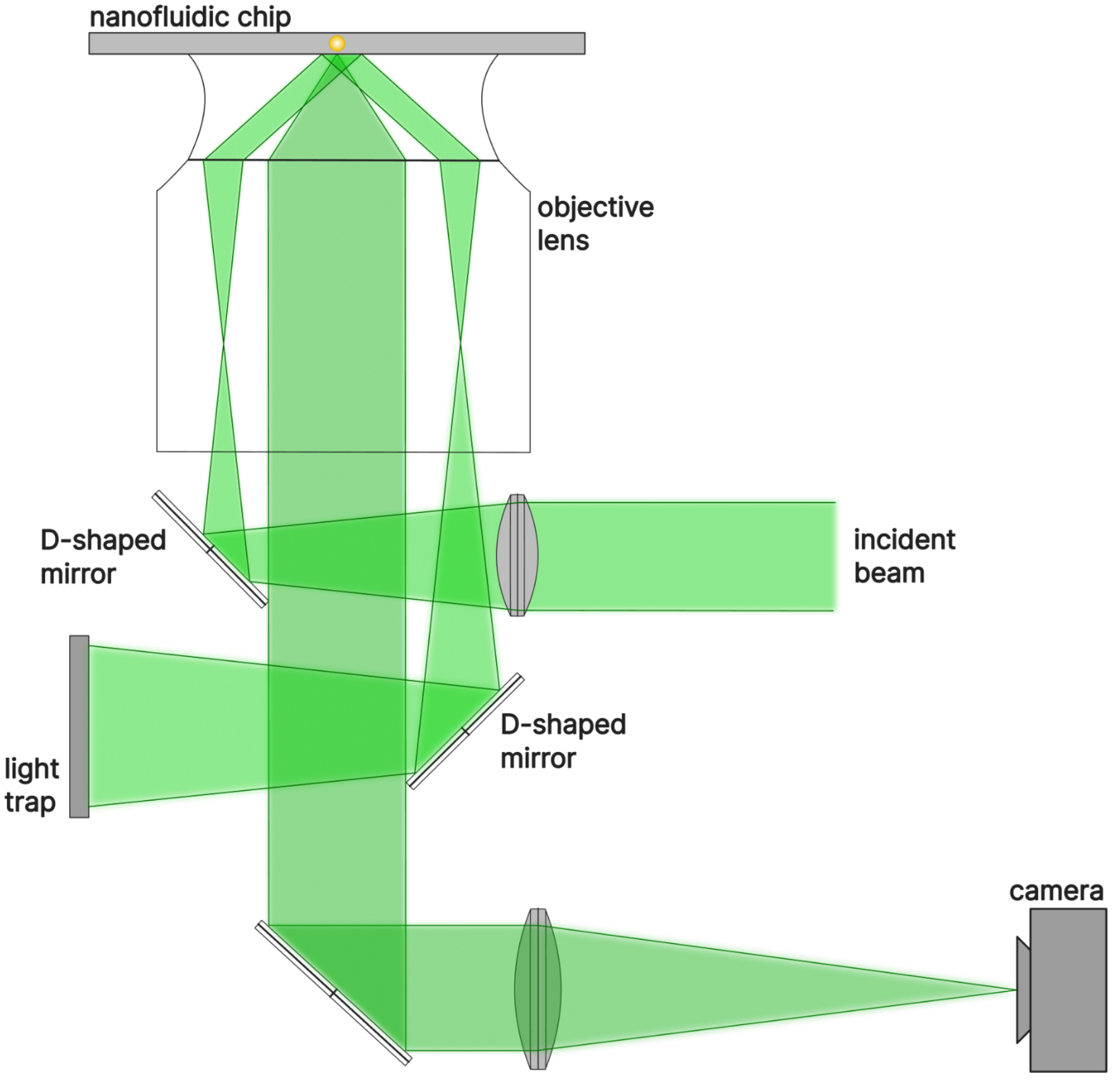

Extended Data Figure 4: Schematic of the optical setup.

## Acknowledgements

The authors acknowledge financial support from the Czech Science Foundation project No. 22-18203S (T.R.) and project No. 21-17847M (N.H., V.P.); from the Programme Johannes Amos Comenius under the Ministry of Education, Youth and Sports of the Czech Republic SENDISO project No. CZ.02.01.01/00/22_008/0004596 (E.E., M.H); from the European Union's Horizon Europe research and innovation programme under the Marie Skłodowska-Curie grant agreement No. 101068106, RAVASI (B.S.); from Czech Academy of Sciences, Premie lumina quaeruntur LQ200402101 (V.P.); from the NWO grant NWA.1292.19.170 (Y.Y., K.T.). Part of this research has been executed at the Dioscuri Centre for Single-molecule Optics infrastructure. The authors thank Adéla Potužníková for her assistance in preparing Extended Data Figure 4. The authors used ChatGPT (OpenAI) to assist with language editing and text refinement. All scientific content, interpretation and conclusions were generated and verified by the authors.

## Author contributions

T.R. and B.S. conceived the study. B.S. developed the theoretical framework and designed the experimental methodology. T.R. and B.S. developed the optical setup and integrated the experimental instrumentation. E.E. developed the image acquisition software. E.E. optimized the experimental instrumentation. M.H. and B.S. developed the NSM data processing pipeline. E.E. performed NSM measurements, collected, processed and analyzed NSM data. N.H. prepared and characterized DNA origami samples under the supervision of V.P.. Y.Y. performed mass photometry measurements under the supervision of K.M.T. B.S. performed FDTD simulations. B.S. supervised the project and acquired funding. T.R., E.E. and B.S. wrote the original draft of the manuscript. N.H., Y.Y., M.H. K.M.T., V.P. contributed to the Materials and Methods section. All authors discussed the results, contributed to the interpretation of the data, reviewed and edited the manuscript, and approved the final version.

# SUPPLEMENTARY INFORMATION

# High-resolution single-molecule mass measurement of megadalton assemblies in solution

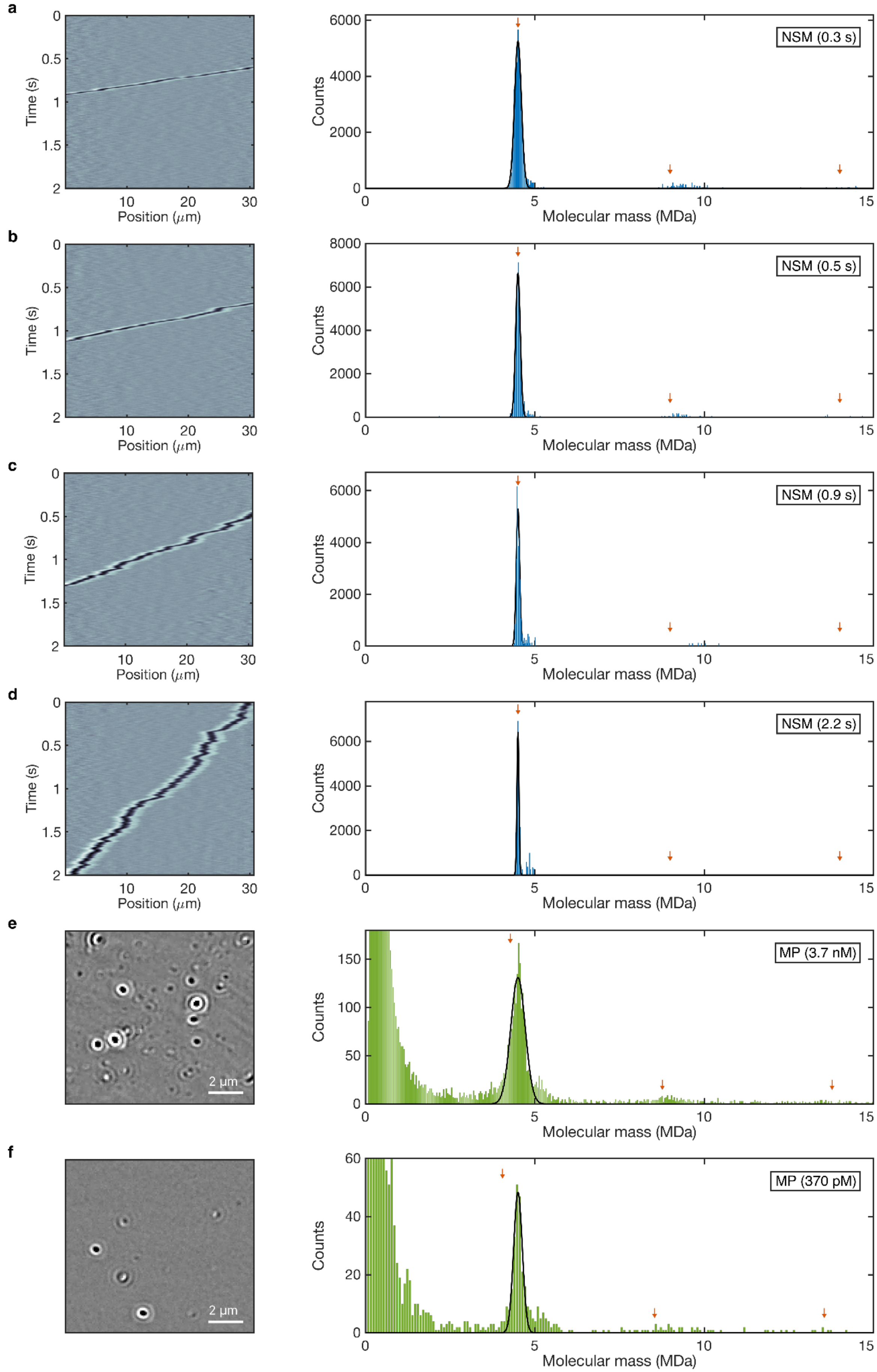
a
b
c
d
e
f
Time (s)
Position (μm)
Counts
Molecular mass (MDa)
NSM (0.3 s)
NSM (0.5 s)
NSM (0.9 s)
NSM (2.2 s)
MP (3.7 nM)
MP (370 pM)
2 μm

SI Figure 1: **(a, b, c, d)** Representative NSM kymographs (left) and mass distributions (right) of individual DNA origami molecules at 60 nM, tracked for 0.3, 0.5, 0.9, and 2.2 s. For each condition, mass distributions were compiled from 914, 663, 219, and 75 trajectories, respectively, acquired under identical illumination conditions **(e, f)** Representative MP differential images (left) and mass distributions (right) for 3.7 nM and 370 pM samples. Arrows in all mass distributions indicate the expected masses of the monomer, dimer, and trimer populations.

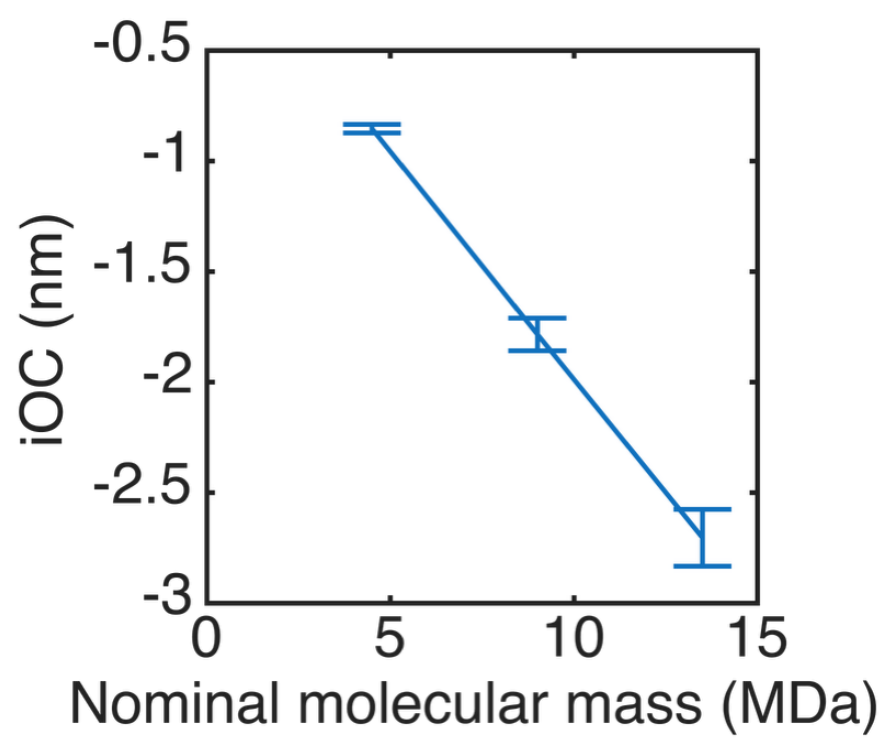

SI Figure 2: Calibration curve obtained from the Gaussian peak centers of monomeric, dimeric, and trimeric DNA origami measured by NSM at a velocity of 96 um/s. The approximately linear relationship between iOC and nominal mass confirms quantitative mass determination; error bars represent Gaussian widths. The monomer, dimer, and trimer populations consisted of 914, 64, and 11 trajectories, respectively.

# Numerical modeling of the interferometric response of DNA origami rods and back-folded conformers

To interpret the experimentally observed subpopulations in interferometric measurements, we modeled the optical response of DNA origami rods and structurally perturbed conformers using finite-difference time-domain (FDTD) simulations (Ansys Lumerical 2025).

DNA origami rods are anisotropic dielectric objects whose optical response depends on their orientation relative to the incident optical field. Their scattering cross-section can therefore be written as $\sigma_{sca} = \sigma_{sca}(\Omega)$, where $\Omega$ denotes the particle orientation with respect to the propagation direction and polarization of the incident light.

The correctly folded extended DNA origami rod was approximated as a homogeneous dielectric cylinder with length $L$ = 200 nm, diameter $d$ = 10nm, and refractive index $n$ = 1.42. To capture structural heterogeneity arising from conformational variability, we introduced a geometrical model of back-folded conformers. In this representation, the rod undergoes partial back-folding into a compact, hairpin-like configuration, while conserving its total contour length. The structure is approximated as two rigid cylindrical segments joined at a junction, with lengths L1 and L2, L1 + L2 = 200 nm. The parameter L1 was varied systematically from 180 to 100 nm, thereby spanning configurations from an unperturbed rod to more compacted, back-folded conformations (SI Fig. 3a). For each geometry, the scattering cross-section was computed for six distinct orientations relative to the illumination direction and polarization (SI Fig. 3b), thereby sampling the anisotropic optical response.

Two spectral regimes were considered: (a) 488 nm, corresponding to the illumination wavelength used in mass photometry (MP), and (b) 500–600 nm, covering the operational range of nanofluidic scattering microscopy (NSM). The results (SI Fig. 4) show a consistent trend across both spectral regimes: the scattering cross-section increases as the structure becomes more compact, i.e., with decreasing $L1$.

For randomly oriented particles under unpolarized (or circularly polarized) illumination, the experimentally relevant observable corresponds to the orientation-averaged scattering cross-section, which was estimated as an average over six sampled arrangements (SI Fig. 3b). Both NSM and surface-based MP measurements can however impose preferential orientation. In NSM, confinement within negatively charged nanochannels is expected to align DNA origami along the channel axis due to steric and electrostatic interactions. In MP, adsorption onto a functionalized glass surface is energetically expected to favor arrangements maximizing contact area, i.e., rods lying flat rather than standing upright. In both cases, the preferential orientation corresponds to situations where the long axis of DNA origami is predominantly perpendicular to the beam propagation direction. To account for these effects, we additionally computed an orientation-restricted average that was estimated as a mean over four corresponding sampled arrangements (SI Fig. 3b).

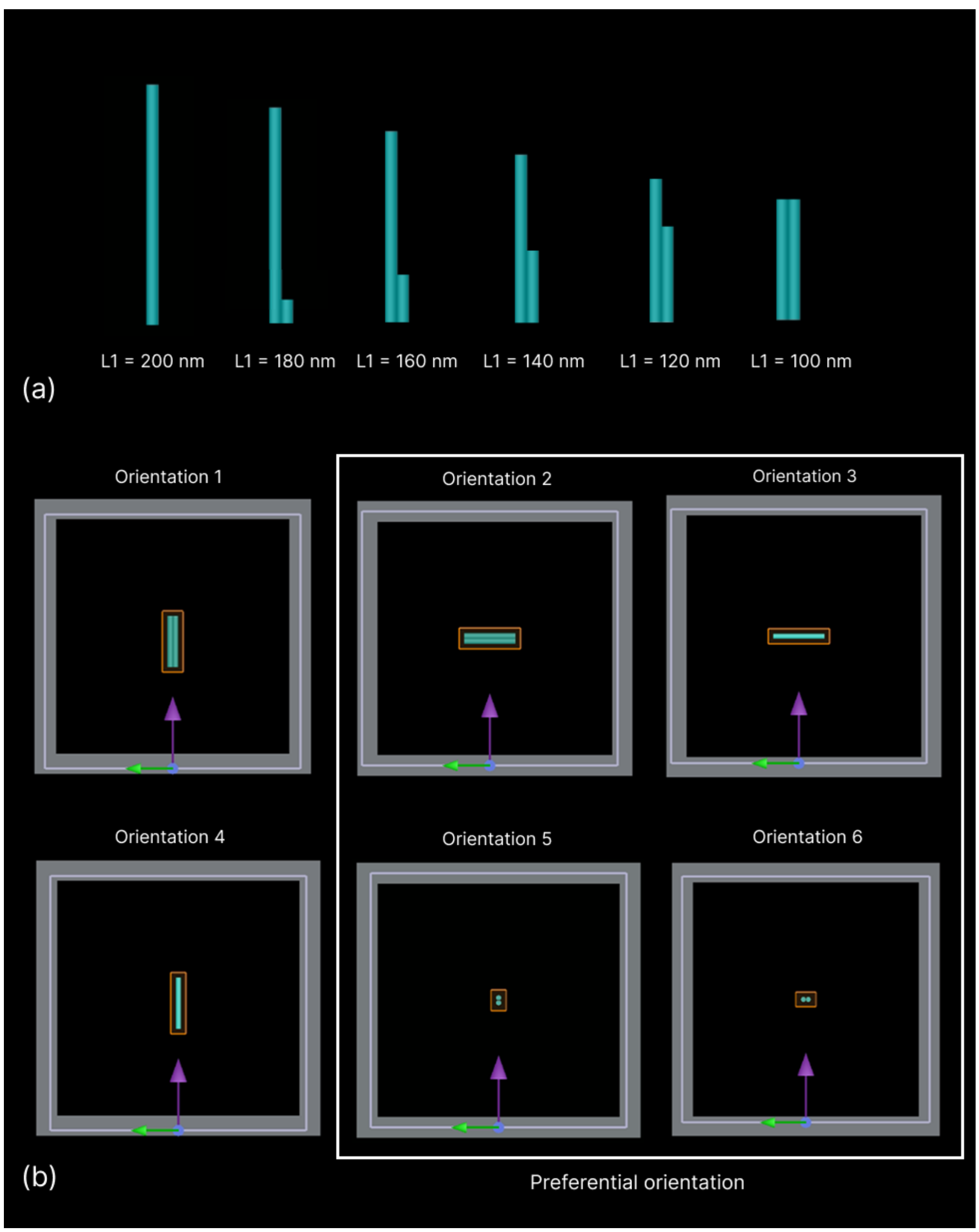

SI Figure 3. Schematic of the simulated structures representing DNA origami rods and back-folded conformers. (a) Series of rods and their structurally perturbed conformers simulated as two cylindrical elements with lengths L1 and L2, L1 + L2 = 200 nm. (b) Six distinct orientations of DNA origami relative to the illumination direction and polarization. The green arrow corresponds to polarization, and the purple arrow corresponds to the beam propagation direction. Preferential orientation corresponds to cases where the long axis of a DNA origami is aligned perpendicular to the beam propagation direction.

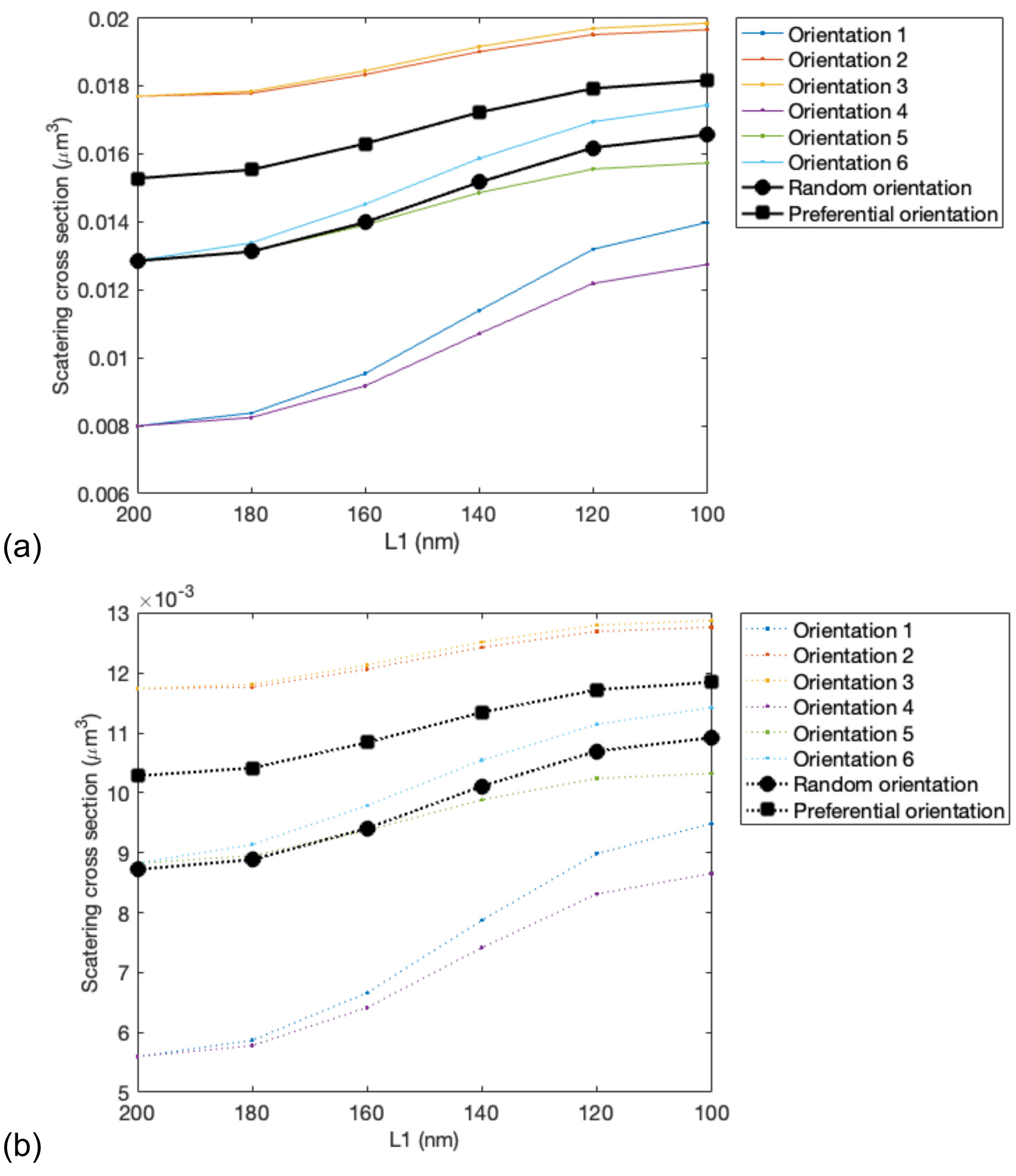

SI Figure 4. Scattering cross section of DNA origami rods and their structurally perturbed conformers of varying L1 for different orientations, and for (a) wavelength 488 nm (corresponding to MP), (b) wavelength range 500-600 nm (corresponding to NSM).

In interferometric detection schemes, the measured signal arises from interference between a reference field and the field scattered by the particle. The detected interferometric contrast can then be written $\approx -2\sqrt{I_m/I_r}$, where $I_r$ is the intensity of the reference wave (reflection from the coverslip in MP, scattering from the nanochannel in NSM), and $I_m$ is the intensity of light scattered by the particle, $I_m \propto \sigma_{sca}$. Therefore, the interferometric optical contrast scales with $\sqrt{\sigma_{sca}}$.

To quantify the expected relative difference of the interferometric optical contrast between DNA origami of different structural compaction, we therefore evaluated $\sqrt{\sigma_{sca}/\sigma_{sca}^0}$, where $\sigma_{sca}^0$ corresponds to the correctly folded extended rod geometry (SI Fig. 4). The simulations predict that structural compaction associated with back-folding leads to an increase in interferometric contrast. This effect is more pronounced at shorter wavelengths (in the MP regime - 488 nm), it is enhanced under random orientation averaging, and reduced when preferential orientation is imposed.

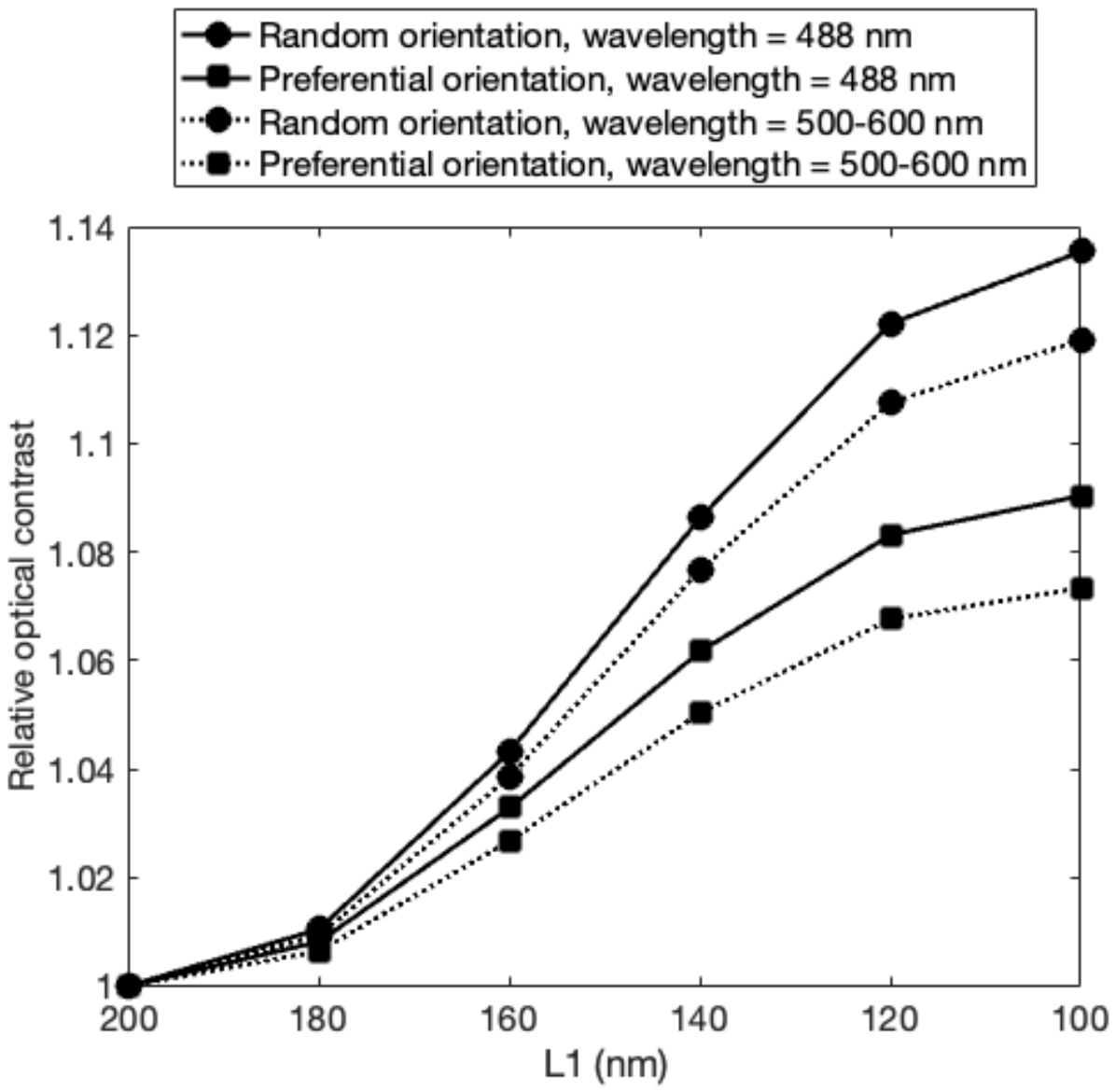

SI Figure 5. Optical interferometric contrast of DNA origami of varying structural compaction (L1) relative to correctly folded extended DNA origami rod.

Taken together, the numerical modeling provides a coherent interpretation of the secondary population observed in NSM and MP. Back-folded DNA origami conformers are predicted to exhibit increased interferometric signal relative to extended rods due to geometry-induced changes in scattering, despite identical molecular mass. The effect is more pronounced in MP, consistent with its shorter wavelength and potentially reduced orientational constraints compared to NSM. These results support the assignment of the additional population to a structurally distinct DNA ensemble rather than a mass-dependent variation.